\documentclass{hep99}
\begin{document}

\newcommand{\sla}{\kern -5.4pt /}
\newcommand{\slalarge}{\kern -10 pt /}
\newcommand{\Dir}{\kern -6.4pt\Big{/}}
\newcommand{\Dirin}{\kern -10.4pt\Big{/}\kern 4.4pt}
\newcommand{\DDir}{\kern -7.6pt\Big{/}}
\newcommand{\DGir}{\kern -6.0pt\Big{/}}

\newcommand{\ra}{\rightarrow}
\newcommand{\be}{\begin{equation}}
\newcommand{\ee}{\end{equation}}
\newcommand{\bea}{\begin{eqnarray}}
\newcommand{\eea}{\end{eqnarray}}
\newcommand{\beanon}{\begin{eqnarray*}}
\newcommand{\eeanon}{\end{eqnarray*}}
\newcommand{\ba}{\begin{array}}
\newcommand{\ea}{\end{array}}
\newcommand{\bd}{\begin{description}}
\newcommand{\ed}{\end{description}}
\newcommand{\bt}{\begin{tabular}}
\newcommand{\et}{\end{tabular}}
\newcommand{\bi}{\begin{itemize}}
\newcommand{\ei}{\end{itemize}}
\newcommand{\ben}{\begin{enumerate}}
\newcommand{\een}{\end{enumerate}}
\newcommand{\bc}{\begin{center}}
\newcommand{\ec}{\end{center}}
\newcommand{\bflr}{\begin{flushright}}
\newcommand{\eflr}{\end{flushright}}
\newcommand{\bfll}{\begin{flushleft}}
\newcommand{\efll}{\end{flushleft}}
\newcommand{\ul}{\underline}
\newcommand{\ol}{\overline}
\newcommand{\dotp}{\!\cdot\!}

\newcommand{\OO}{{\it O}}
\newcommand{\gi}{{\it gi}}
\newcommand{\GI}{{\it GI}}

\parindent 0cm
\newcommand{\parno}{\par\noindent}

\newcommand{\ar}{\rightarrow}
\newcommand{\vsk}{\vskip 10 pt\noindent}
\newcommand{\hsk}{\hskip 10 pt\noindent}
\newcommand{\vskk}{\vskip .5cm\noindent}
\newcommand{\hskk}{\hskip .5cm\noindent}

\newcommand{\ph}{{\tt PHACT\ }}
\newcommand{\wph}{{\tt WPHACT\ }}
\newcommand{\sph}{{\tt SIXPHACT\ }}

\def\epem{\ifmmode{e^+ e^-} \else{$e^+ e^-$} \fi}
\newcommand{\eeww}{$e^+e^-\rightarrow W^+ W^-$}
\newcommand{\qqQQ}{$q_1\bar q_2 Q_3\bar Q_4$}
\newcommand{\eeqqQQ}{$e^+e^-\rightarrow q_1\bar q_2 Q_3\bar Q_4$}
\newcommand{\eewwqqqq}{$e^+e^-\rightarrow W^+ W^-\ar q\bar q Q\bar Q$}
\newcommand{\eeqqgg}{$e^+e^-\rightarrow q\bar q gg$}
\newcommand{\eeqqqq}{$e^+e^-\rightarrow q\bar q Q\bar Q$}
\newcommand{\eewwjjjj}{$e^+e^-\rightarrow W^+ W^-\rightarrow 4~{\rm{jet}}$}
\newcommand{\eeqqggjjjj}{$e^+e^-\rightarrow q\bar 
q gg\rightarrow 4~{\rm{jet}}$}
\newcommand{\eeqqqqjjjj}{$e^+e^-\rightarrow q\bar q Q\bar Q\rightarrow
4~{\rm{jet}}$}
\newcommand{\eejjjj}{$e^+e^-\rightarrow 4~{\rm{jet}}$}
\newcommand{\jjjj}{$4~{\rm{jet}}$}
\newcommand{\qqbar}{$q\bar q$}
\newcommand{\ww}{$W^+W^-$}
\newcommand{\sm}{${\cal {SM}}$}
\newcommand{\wwqqqq}{$W^+ W^-\ar q\bar q Q\bar Q$}
\newcommand{\qqgg}{$q\bar q gg$}
\newcommand{\qqqq}{$q\bar q Q\bar Q$}

\newcommand{\rd}{{\mathrm{d}}}

\def\cpc #1 #2 #3 {{\em Comp. Phys. Commun.} {\bf#1} (#2) #3 } 
\def\np #1 #2 #3 {{\em Nucl.~Phys.} {\bf#1} (#2) #3 }
\def\pl #1 #2 #3 {{\em Phys.~Lett.} {\bf#1} (#2) #3 }
\def\npps #1 #2 #3 {{\em Nucl. Phys. Proc. Suppl.} {\bf#1} (#2) #3 }
\def\zp #1 #2 #3 {{\em Z.~Phys.} {\bf#1} (#2) #3 }

\begin{titlepage}
\setcounter{page}{0}
\topmargin 2mm
\begin{flushright}
{\large DFTT 59/99}\\
{\rm October 1999\hspace*{.5 truecm}}\\
\end{flushright}

\vspace*{3cm}
\begin{center}
{\Large \bf \noindent LEP2 4f review 
\footnote{ Work supported in part by Ministero
dell' Universit\`a e della Ricerca Scientifica\\ 
e-mail: ballestrero@to.infn.it}}
\\[1.5cm]

{\large  Alessandro Ballestrero 
   }\\[.3 cm]

{\it I.N.F.N., Sezione di Torino and \\
 Dipartimento di Fisica Teorica, Universit\`a di Torino}\\
{\it v. Giuria 1, 10125 Torino, Italy.}

\vspace*{4cm}

\centerline{\bf ABSTRACT}
\vsk
{\normalsize\noindent 
We  review some recent results and open problems on four fermion 
physics at  LEP2 and beyond.
} 

\vskip 5cm
{Invited talk given at the\\\smallskip {\em EPS Conference on High
Energy Physics}\\ Tampere, Finland,
July 1999}\\
\smallskip
{\em to be published in the proceedings}\\

\end{center}
\end{titlepage}


\title{LEP2 4f review }

\author{Alessandro Ballestrero}
%

\address{INFN and Dip. Fisica Teorica - Via Giuria 1 - 10125 Torino - Italy \\[3pt]
E-mails: {\tt ballestrero@to.infn.it}}

\abstract{ We  review some recent results and open problems on four fermion 
physics at  LEP2 and beyond.
} 

\maketitle


\section{Introduction}
LEP2 has opened the era of four fermion physics.  These processes 
have shown to be essential for deeper insight in the  SM and 
their precise knowledge is fundamental for limits and searches
of signals of New Physics.

Important experimental results in this field have already been achieved
(see other talks in this session). 

On the theory side, a strong effort has led in recent years to
quite a number of dedicated 4f MC generators.
Among them, some are capable to compute (with different
methods, features, approximations etc.) the whole set of 4f final states:
{\tt EXCALIBUR \cite {exc}}, which has been the first complete code, 
{\tt ALPHA\cite{alp}, COMPHEP\cite{com}, KORALW\cite{kor}, GRC4F\cite{grc}, 
WPHACT\cite{wph}, WTO\cite{wto}, WWGENPV\cite{wwg}}.
The only 4f code to use a semianalytical approach is {\tt GENTLE}\cite{gen}, 
which can compute all  final states with no  identical particles, 
$e$'s or $\nu_e$'s.

Good technical agreement among these generators has been tested during the 
first LEP2 workshop\cite{yr}.

Despite the very high technical precision of the codes, the theoretical 
uncertainty for the 4f processes is normally \OO (1$\div$2\%).
But of course it  depends very much on final states and on cuts and
for some configurations (eg. arbitrarily low invariant masses and angles) 
we do not even have a safe estimate.

Already with the present LEP2 luminosity ({ L $\approx$ 1200 pb$^{-1}$})
many statistical errors are comparable or even smaller than theoretical ones.
Therefore, in view of the foreseen total luminosity and of the many different 
physical processes explored, new requests come from the experimental community
for a higher precision. Also the Linear Collider (LC) will pose further 
severe requests on the reliability of  4f computations.
In the following, after recalling gauge invariance issues, we will give a  
short review of  recent developments in this sense
and of some relevant open problems.

\section { The gauge invariance issue} 

As it is well known, perturbative  calculations should be 
 gauge invariant (\gi) , otherwise the 
result can depend in an uncontrolled way on the chosen gauge.

4f calculations and their radiative corrections,  involve in general three 
types  of possible gauge invariance (\GI) problems related to: 
the use of an incomplete set of diagrams,  
the   non \GI\ of some initial (ISR) or final (FSR) 
state radiation treatment,  
the fact that one has to deal with unstable particles.

As an example of the first type, we may recall that
the double resonant diagrams for WW (CCO3)  or ZZ (NCO2)
production are the dominant contributions
but they are not \gi\ as other diagrams (single or non resonant ones) 
contribute to the same final state. The latter are however \OO($\Gamma\over M$)
 and suppressed for invariant mass cuts near M.

For \GI\ and ISR/FSR, let us recall that to
 have a \gi\ quantity, one must consider the emission of a photon 
from all possible parts of a charged line. So for instance in WW production
diagrams one cannot just consider the emission from incoming or outgoing 
fermions, as the charged line passes through the W which can also emit. 
As a consequence, ISR and FSR alone are \gi\ only in LL approximation,
and this implies for instance 
that accounting for $p_T$ $\gamma$'s via ISR/FSR might in
principle be dangerous.

The presence of unstable particles can easily lead to violating gauge 
invariance :
intermediate W or Z give rise at tree level to 
poles $1\over {p^2-M^2}$.
This is cured by introducing finite decay width:
${ 1\over {p^2-M^2}}  \ar  {1\over {p^2-M^2+i\,\Pi}} $, with
$\Pi$=$\Gamma$M (fixed width)  or  $\Pi$=$\Gamma$${p^2}\over M$ (running
width).
It corresponds to the Dyson resummation of self energy graphs. In this way 
 part of higher order corrections are accounted for and the procedure
 breaks \GI, which works order by order in 
perturbation theory. In general this results in negligible effects 
 \OO($\Gamma\over M$), but they can be dramatically enhanced by
gauge cancellations or small scales, as it happens for instance in so called
 single W processes.
Different possible solutions are known:
\bd
\item {-} Fixed width. It makes use of fixed width also in t channel 
          Vector Boson propagators. 
\item {-} Overall scheme.  All tree level diagrams 
      (not only the resonant ones) are multiplied by factors 
       ${{p^2-M^2}\over {p^2-M^2+i\,\Gamma M}}$. 
\item {-} Pole scheme\cite{ps}.
        Expand the amplitudes in $\Gamma\over M$ around the complex poles. 
        Each order of the expansion will be separately \gi.
\item {-} Fermion loop (FL) scheme\cite{fl}.  
        Use Dyson resummed resonant propagator with 
         (at least) all other one loop corrections necessary to satisfy
	the Ward identities. 
        This is considered to be the most satisfactory
        from a theoretical point of view
\ed

\section { Radiative corrections - DPA} 

Full EW RC to on shell W production and decay 
 are known.
Complete tree level for 4f (+ ISR/FSR, Coulomb, hadronization, PS, 
hadronization, ..) are
implemented in MC event generators.
Complete EW one loop corrections to full ME are incredibly complicated
 and at  present not available.

A consistent, gauge invariant way of evaluating the most relevant corrections
to ``WW (or ZZ) like`` full processes has been studied in 
recent years (DPA)\cite{ddr} \cite{bbc}
and completed recently\cite{bbc}.
It corresponds to the above mentioned Pole scheme and it is referred to
as  Double Pole Approximation 
With these corrections $\approx$.5\% theoretical precision will 
hopefully be achieved in WW  cross sections and similar methods can be extended
to ZZ.

In DPA one  expands the  amplitude around complex poles of unstable W's.
Only double resonant terms are retained in tree level and \OO($\alpha$)
corrections.
It corresponds to an expansion both in $\alpha$ and ${\Gamma_W}\over {M_W}$
Terms \OO(1), \OO($\alpha$), \OO(${\Gamma_W}\over {M_W}$)
are kept;  
\OO($\alpha{\Gamma_W}\over {M_W}$) is neglected.
Resonant propagators can be safely be resummed and the various orders
are separately \gi.
The method is reliable only above threshold and for cuts on invariant masses
not far from the poles.

LO cross section $d\sigma^0$ for a certain final state 
is splitted in

\[ d\sigma^0 = d\sigma^0_{DPA} + (d\sigma^0-d\sigma^0_{DPA}) \]

$d\sigma^0_{DPA}$ consists of a \gi\ approximation of CCO3 in which
production and decay parts are computed on shell ($p_i^2 \ar M^2$),
spin correlations are exactly accounted for
and the resummed W propagators maintain the correct dependence on 
      off shell masses.


\begin{figure}
  \begin{center}
  \unitlength 1cm
  \vspace*{-.1cm}
  \begin{picture}(5.,5.)
  \put(-1.9,-3.8){\includegraphics{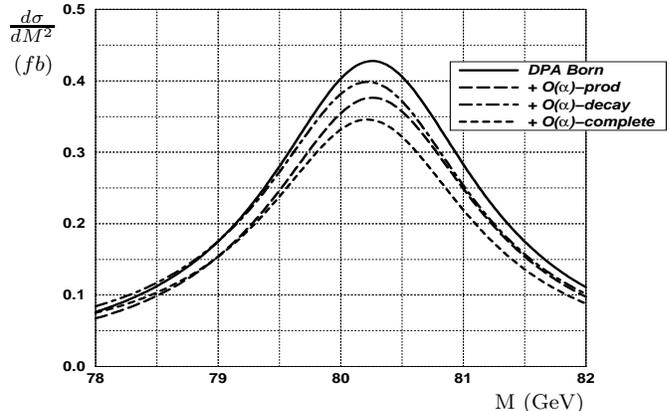}}
  \put (5.,-.5){\footnotesize M (GeV)}
 \put (-1.6,4.5){ $\frac{d \sigma}{dM^2}$}
 \put (-1.4,4.){\footnotesize $(fb)$}
  \end{picture}
  \end{center}
\caption {Invariant mass distribution for the $\mu^+\nu_\mu\tau\bar\nu_tau$ 
final state
(from ref.~\cite{bbc})}
\end{figure}

DPA real and virtual \OO($\alpha$) corrections $\delta_{DPA}$ are then
computed.

They can be divided in factorizable and non factorizable.
The first ones apply separately to production and decay parts, while the
second account for interference effects between different parts
and are in general of limited numerical relevance.

The final result is

\[ d\sigma = d\sigma^0_{DPA}(1+\delta_{DPA}) + (d\sigma^0-d\sigma^0_{DPA}) \]

In fig~.1 is reported the W invariant mass spectrum computed in DPA, where the
complete calculation show a sizeable shift.
Exact complete DPA are at present  not yet implemented in a MC.

%




\section { 4f + $\gamma$}

$\epem \ar 4f \gamma$ processes with both 
  visible and invisible $\gamma$ are important. 
Even if $4f+$ visible $\gamma$ events are not as abundant at LEP2 as
they will be at  LC, they
involve triple and quartic gauge coupling and  are relevant for W mass 
measurement and for  searches. 
Moreover,
$\epem \ar 4f \gamma$  calculations are a fundamental step towards having a 
description of radiation that goes beyond 
$p_T$ ISR/FSR and its \GI\ problems.
The goal is to match them with radiative corrections and collinear radiation.

Calculations with massive fermions + $\gamma$ have already been performed for 
CC10 ($\mu \nu u \bar d \gamma $) and CC20 ($e \nu u \bar d \gamma$)
\cite{alg} processes.
Recently a new massive computation of ($\mu \nu u \bar d \gamma $) has been 
completed \cite{jeg}, which  makes use of 
multichannel MC  techniques.

Results from a complete $4f+\gamma$ MC ({\tt RACOONWW}) have appeared
this year\cite{ddrw}.
 This is at present the only code
which can account for all $4f$ and $4f + \gamma$ final states with all 
tree level   Feynman diagrams. Other important {\tt RACOONWW} features are:
the multichannel integration, the non linear gauge fixing 
which  simplifies the calculations 
and the gauge restoring scheme which preserves
SU(2)xU(1) using everywhere  complex boson masses  
         ($M^2\ar M^2-i\,\Gamma M$). The code assumes 
 massless fermions, therefore collinear and soft regions have to be excluded
        with cuts and so far it has 
 no ISR/FSR and interface to hadronization.

An interesting example of photon spectra calculation by {\tt RACOONWW}
is reported in fig.~2

\begin{figure}
  \unitlength 1cm
  \vspace*{0.cm}

\centerline{
\setlength{\unitlength}{.55 cm}
\begin{picture}(14.5,6.3)
\put(3.6,14){\includegraphics{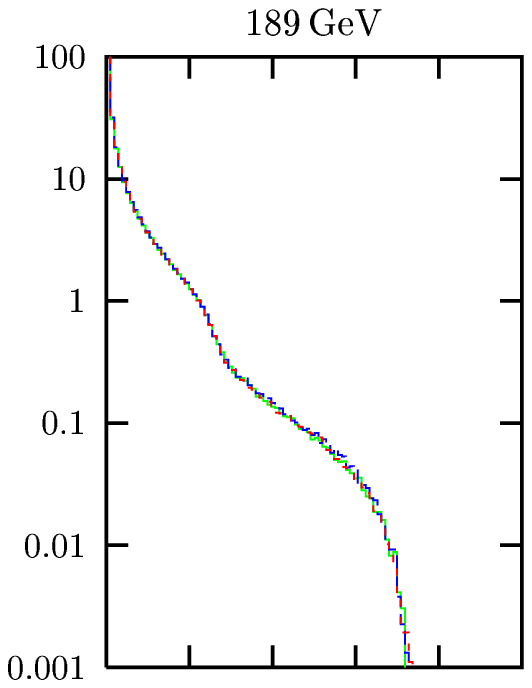}}
\put(7.9,14){\includegraphics{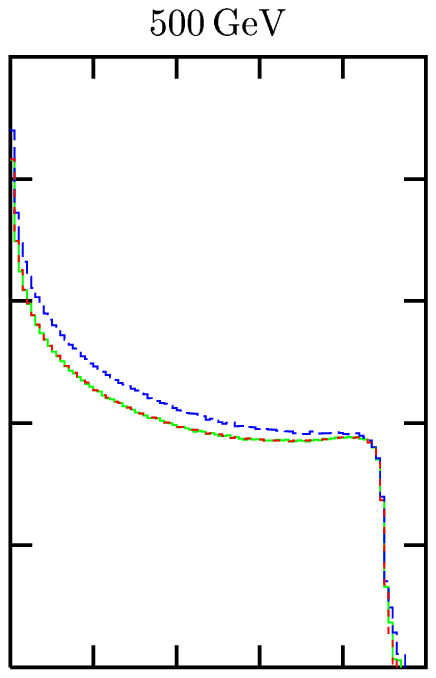}}
\put(12.2,14){\includegraphics{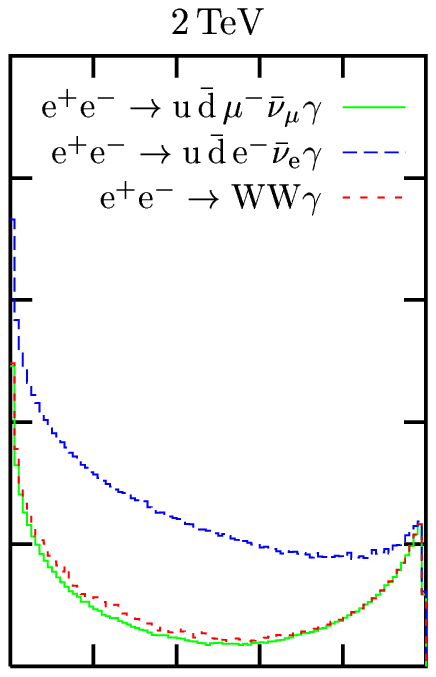}}
\put(-.5,5.5){\makebox(1,1)[c]{\footnotesize $\frac{\rd \sigma}{\rd E_\gamma}$}}
\put(-.5,-.5){\makebox(1,1)[c]{\footnotesize $
                                 \frac{fb} {GeV}$}}
\end{picture}
}
\centerline{
\setlength{\unitlength}{.55 cm}
\begin{picture}(14.5,6.3)
\put(3.6,14){\includegraphics{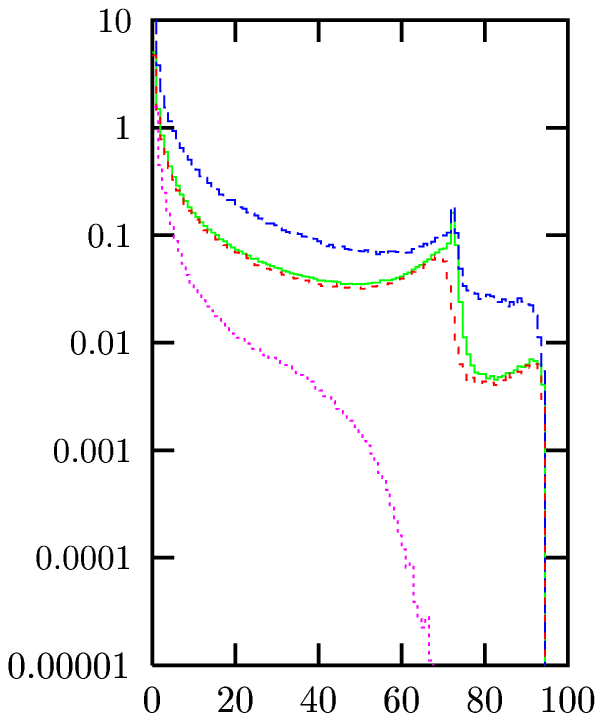}}
\put(7.9,14){\includegraphics{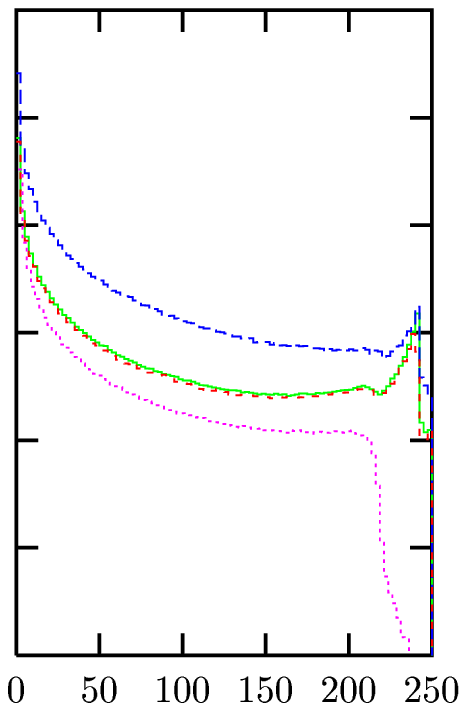}}
\put(12.2,14){\includegraphics{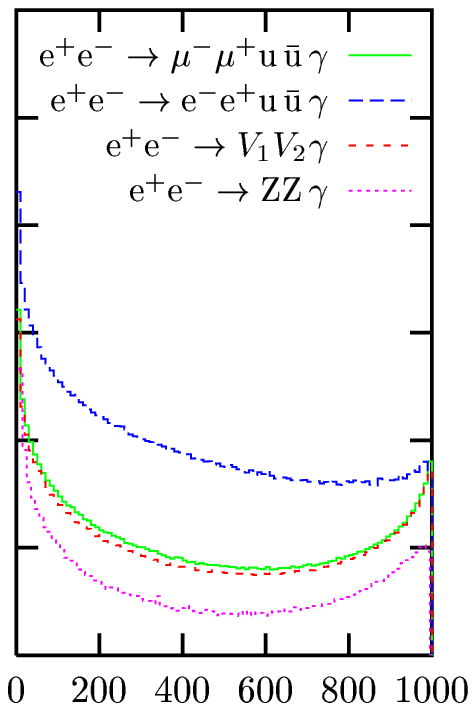}}
\put(7.2,-0.7){\makebox(1,1)[cc]{{\footnotesize $E_\gamma/ GeV$}}}
\end{picture}
}
\caption {$d\sigma/dE_\gamma$ (fb/GeV) photon spectra from the 
complete calculation and from triple gauge boson production (from ref.\cite{ddrw})}
\end{figure}
 

\section {\bf Single W} 

The charged current  processes with an electron and its  neutrino 
in the final state 
($e\bar\nu f_1 \bar f_2$)  have,
besides the usual diagrams of  the corresponding process 
\hsk $\mu\bar\nu f_1 \bar f_2$, all 
 diagrams obtained exchanging  the incoming $e^+$ with the outgoing
$e^-$. These contributions become dominant for $\theta_e \ar 0$
 for the presence of  the t-channel 
$\gamma$ propagator. These four fermion events with
  $e$ lost in the pipe are often referred to as single W,
and they are relevant for triple gauge studies and as background to searches.

The t channel $\gamma$ propagator diverges at $\theta_e=0$ in the
$m_e \ar 0$ limit,
so fermion masses have to be exactly accounted for.
Moreover, the ${1}\over {t^2}$ behaviour is reduced to  ${1}\over {t}$ by 
gauge cancellations. This implies that even a little violation of gauge
conservation can have dramatic effects and the use of some 
gauge conserving scheme is unavoidable.
Finally, the  phase space integration has to be studied carefully for
the 
logarithmic behaviour in $m_e$ and also in other masses for 
low $m(f_1\bar f_2)$ (multiperipheral diagrams).

Two strategies have been used so far: Improved Weiszacker Williams 
(IWW)\cite{iww} and completely massive codes.

In the first case one separates the 4 t-channel photon diagrams, evaluates
them analytically in equivalent photon approximation with exact  
dependence on all masses, and then adds
the rest of diagrams + interference in the massless approximation.
An example of results obtained with this approach is given by the m(ud)
mass distribution of fig.~3.

In the fully massive MC numerical approach { \tt COMPHEP, GRC4F, KORALW} and
{\tt WPHACT}
have recently compared their results and found a very good agreement reported
in fig.~4. An  agreement of the order of .5\% has been obtained   by
{ \tt COMPHEP, GRC4F, WPHACT} also at 800 GeV.
It has to be noticed that the version of {\tt WPHACT} used in these comparison
is a new one in which all completely massive matrix elements have been
added, so that one can choose between these and the fastest massless ones.
In fig.~4 are reported also the results obtained by the massless codes 
{\tt EXCALIBUR} and {\tt ERATO}\cite{era} using
a fictitious minimal angle to avoid the divergence for $m_e=0$. 

\begin{figure}
  \begin{center}
  \unitlength 1cm
  \vspace*{.5cm}
  \begin{picture}(5.,5.)
  \put(-1.8,-1.7){\includegraphics{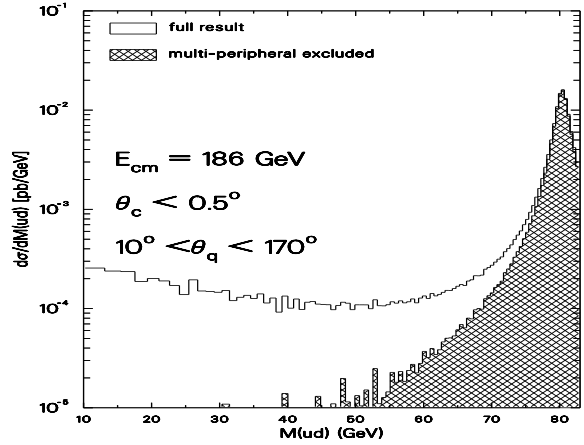}}
  \end{picture}
  \end{center}
\caption{Invariant mass m(ud) distribution for the process $e^+ e^- \ar e^- \nu u \bar d$ computed by {\tt WTO}}
\end{figure}


In view of the relevance of using a gauge restoring scheme in these 
computations, Massive FL (Im part) scheme has recently been implemented 
\cite{abm}. The comparison among this scheme and  the others reported
in fig.~5 shows a substantial independence from the \gi\ scheme.



\begin{figure}
  \begin{center}
  \vspace*{-1.5cm}
  \unitlength 1cm
  \begin{picture}(5.,5.1)
  \put(-2.5,-1.8){\includegraphics{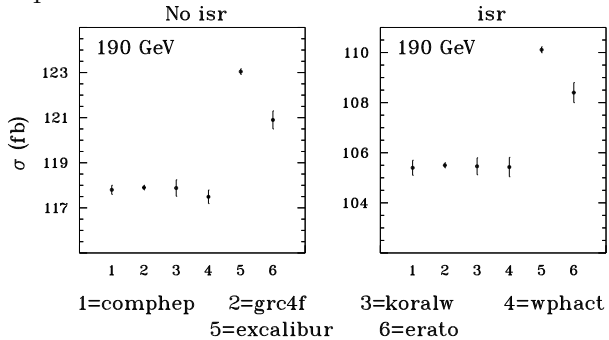}}
  \end{picture}
  \end{center}
\caption {Comparison on the total cross section for $e^+ e^- \ar e^- \nu u 
 \bar d$ with $|cos \theta_e| > .997$, $m(ud) > 5$ GeV, $E_{u,d} > 3$ GeV}
\end{figure}

\begin{figure}
  \begin{center}
  \vspace*{-1.cm}
  \unitlength 1cm
  \begin{picture}(5.,5.)
  \put(-2.5,-2){\includegraphics{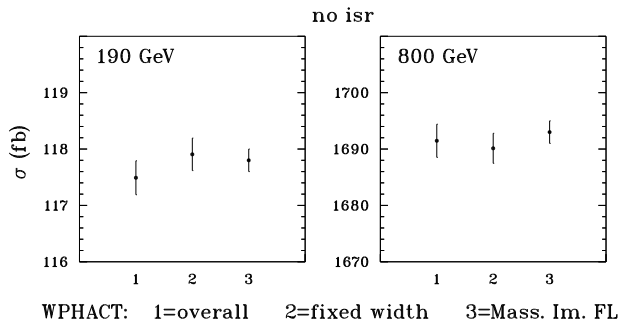}}
  \end{picture}
  \end{center}
\caption { Total cross sections for different gauge restoring
schemes computed by {\tt WPHACT} for the same process and cuts as in fig.~4}
\end{figure}


\section { Conclusions } 

Recent developments show that 4f physics is
at present leaving the percent era.
The new precision requires better knowledge of RC to full 4f final states.
 Important results in this sense  
will probably  allow a sensible diminution of the theoretical error
for  WW or double resonant physics.
However, also other channels, processes and  selection cuts are under active
experimental investigation and in many cases 
 the actual theoretical precision is much worse than
1\%. 
In  LEP2 MC workshop, 4f subgroup, these topics are at present under discussion.

Some of the  main open problems which still need a definitive answer regard
\bi 
\item  Inclusion of photons and ISR/FSR:\\
        beyond $p_T$ ISR/FSR; 
        radiative and non radiative events;
       how to match the two regimes.
\item  4F contribution to monojets
\item  ``Single W'' ($e\nu 2q$ with lost e) for 2q's with low 
          invariant masses: resolved photons. 
\item  Single Z  
\item  Gamma Gamma physics with 4f and relationship
 with dedicated gamma gamma MC
\item  t channels and ISR/FSR
\item  Radiative corrections:\\
      DPA for ZZ;
      for MIXED processes;
      for non WW or ZZ physics;
       complete \OO ($\alpha$);
         LC and radiative corrections.
\ei

Needless to say, their solution is in some case far from easy but
one must realize that these investigations,  important for LEP2, will become 
essential in the future for all physics studies at Linear Collider.

\end{document}